\documentclass[twocolumn,showpacs,preprintnumbers,amsmath,amssymb,floatfix,superscriptaddress]{revtex4-1}
\usepackage{lineno}
\usepackage[utf8]{inputenc}
\usepackage{amsmath}
\usepackage{amssymb}
\usepackage{graphicx}
\usepackage{epstopdf}
\usepackage{graphicx}
\usepackage{color}
\usepackage{epsfig}
\usepackage{hyperref}
\usepackage{easyReview}
\DeclareUnicodeCharacter{0416}{}
\DeclareUnicodeCharacter{0443}{}
\DeclareUnicodeCharacter{0440}{}
\DeclareUnicodeCharacter{043D}{}
\DeclareUnicodeCharacter{0430}{}
\DeclareUnicodeCharacter{043B}{}
\DeclareUnicodeCharacter{044D}{}
\DeclareUnicodeCharacter{043A}{}
\DeclareUnicodeCharacter{0441}{}
\DeclareUnicodeCharacter{043F}{}
\DeclareUnicodeCharacter{0435}{}
\DeclareUnicodeCharacter{0438}{}
\DeclareUnicodeCharacter{043C}{}
\DeclareUnicodeCharacter{0442}{}
\DeclareUnicodeCharacter{044C}{}
\DeclareUnicodeCharacter{043E}{}
\DeclareUnicodeCharacter{0439}{}
\DeclareUnicodeCharacter{0447}{}
\DeclareUnicodeCharacter{0444}{}
\DeclareUnicodeCharacter{0437}{}

\begin{document}

\title{Estimation of conserved charges for a one dimensional system with inhomogeneous hopping}
\author{Triparna Mondal} 
\affiliation{Indian Institute of Science Education and Research Kolkata, Mohanpur, 741246 India}

\date{\today}

\begin{abstract}
Quantum integrability in a system is characterized by a large number of conserved charges. However, finding the conserved charges in a generic quantum system is highly challenging. The integrable matrix theory provides a unified framework to obtain the conserved charges in a certain class of systems.
We demonstrate this framework in case of a single-particle system on a 1D finite-sized lattice with inhomogeneous nearest neighbor hopping to study the statistical properties of the system across its chaotic--integrable crossover from the perspective of the conserved charges.
The eigenspectrum of our random matrix model is studied first. We then estimate the conserved charges and find their properties throughout the chaotic to integrable transition of the system. We calculate the number of conserved charges across this crossover and observe that it is nearly equal to the size of the system at its integrable limit. Our result suggests that the number of conserved charges, estimated using the integrable matrix theory, can be a measure of quantum integrability.

\end{abstract}
\maketitle

\section{Introduction}
\label{s1}

A random matrix ensemble is defined as a collection of Hamiltonian matrices whose elements are random numbers, sampled from a specific distribution. The spectral statistics of such ensembles typically exhibit signatures of quantum chaos, characterized by correlated energy levels and strong level repulsion \cite{haake2010, mehta1991}. On the other hand, Poissonian spectral statistics are often associated with quantum integrability in random matrix ensembles, where energy levels are uncorrelated \cite{berry1977, relano2004}. Similar Poissonian statistics may also arise in many-body localized phases \cite{pal2010, andraschko2014}.
1D Anderson ensemble is one of such disordered ensembles which shows localization in its eigenfunctions leading to its eigenvalues obeying Poisson statistics. Strong random onsite disorder along with homogeneous nearest neighbor hopping in the Hamiltonian of this system gives rise to its integrable behavior \cite{anderson1958, evers2008}.

Quantum integrability for Hamiltonian matrices beyond Poisson statistics is largely under-explored. One of the important measures of integrability in various systems could be the existence of parameter-dependent conserved charges \cite{grabowski1995,grosse1989,luscher1976,stefan1995}.
In Ref. \cite{modak2016}, conserved charges were introduced to explain localization in 1D Anderson-type ensemble.  The foundation of this kind of analysis of integrability in terms of conserved charges is the integrable matrix theory (IMT) \cite{shastry2016a}; since IMT is established based on all the the criteria of quantum integrability for a finite-sized system \cite{shastry2005, shastry2013}. Those criteria mainly include: (a) exact solution of the spectra, (b) Poisson level statistics, (c) level crossing of same symmetry in parameter-dependent Hamiltonian \cite{shastry2005}.

The integrability in the eigenspectrum along with the localization in the eigenstates of a 1D Anderson ensemble is expected to be affected if inhomogeneity is introduced in the hopping parameters of their Hamiltonian. This leads to an important question: how satisfactory are the conserved charges as a measure of quantum integrability for 1D systems where hopping is no longer homogeneous?

Here, we try to formulate a 1D system where the nearest neighbour hopping parameters are random. To introduce randomness in the hopping, we look for a simple and popular ensemble. The $\beta$-ensemble is well-known for the
simple structure of the matrix representation of its Hamiltonian \cite{edelman2002}. The Hamiltonian of this ensemble in its matrix form is similar to that of the 1D Anderson ensemble except the hopping terms here considered from a $\chi$-distribution of certain degrees of freedom. An appropriate parameter based tuning of those degrees of freedom resulted to a non-ergodic-extended phase in the crossover from chaotic to integrable limit of its spectrum \cite{adway2022}. In this phase, the spectral statistics deviate from the Wigner-Dyson statistics \footnote{The chaotic statistics in random matrix theory are often termed as Wigner-Dyson statistics from the names of the scientists who discovered this theory \cite{haake2010, mehta1991}} but do not reach the integrable limit. Thus, it is quite intriguing to study such degrees of freedom based transition in terms of conserved charges. 
In this paper, we focus on the formulation of a 1D ensemble with inhomogeneous nearest neighbour hopping, where inhomogeneous hopping terms are sampled from a $\chi$-distribution with certain degrees of freedom. We estimate the conserved charges for this novel system using IMT
and study their properties across the chaotic--integrable crossover. We compute the number of conserved charges spanning the crossover, as it is considered to be an essential measure of the strength of quantum integrability in a system. A completely integrable system of finite size is expected to possess a number of conserved charges equal to its size \cite{shastry2005}. This criterion has been verified for our system at its completely integrable limit.

The manuscript is organised as follows. In section \ref{s2}, we briefly introduce IMT and following that, the formulation of conserved charges
 is discussed. Section \ref{s3} formulates the model with 1D inhomogeneous hopping. Although, the inhomogeneity is achieved by introducing $\chi$-distributed random numbers, our model is far from the well-known $\beta$-ensemble. The reason why IMT does not directly lead to the $\beta$-ensemble is also discussed here. The statistical properties of our novel ensemble is studied in the section \ref{s4}. The latter part of this section \ref{s53} discusses how the spectral as well as eigenstate properties of our system differ from those of the $\beta$-ensemble. Section \ref{s5} is dedicated for the estimation of the conserved charges for our model. Here, the properties and the numbers of conserved charges are thoroughly studied across the chaotic--integrable crossover of the spectrum.
In the final section \ref{s6}, we summarize the obtained results and conclude.

\section{Conserved charges from IMT}
\label{s2}

In finite dimensional context, IMT was introduced by satisfying all the criteria for general quantum integrability \cite{shastry2005}. This theory provides a recipe to study quantum integrable systems and their characteristics. In this framework, the Hamiltonian is considered to be of the form: 
\begin{equation}
H=V+xT~.
\label{H}
\end{equation}
Here $H,V$ and $T$ are $N\times N$ Hermitian matrices and $x$ is a real parameter, tuning which a chaotic--integrable transition can be achieved \cite{owusu2013, shastry2016}.
At $x=0,~H$ is completely chaotic with its eigenvalues satisfying Wigner-Dyson statistics. This implies that at $x=0$, if $V$ is written in its eigenbasis, it is a diagonal matrix with elements distributed according to the Wigner semicircle law. In the same eigenbasis of $V$, for a non-zero $x$, if the matrix $T$ is constructed as a function of eigenvalues and eigenvectors of $V$ (detailed procedure discussed in Refs.~\cite{shastry2016a, shastry2016}), the system moves towards its integrable limit with increasing $x$ and at $x=1$, it becomes completely integrable.  

For integrable Hamiltonian $H$, there exist conserved charges $H^i$  satisfying equations of the form same as Eq.~\ref{H}, \textit{i.e.} $H^i=V^i+xT^i$. All $H^i,V^i$ and $T^i$ are again Hermitian matrices of size $N$ with $V^i$ being diagonal and $T^i$ being defined in a manner analogous to $T$ in the same basis. Conserved charges $H^i$ satisfy the following commutations,
\begin{equation}
[H(x),H^i(x)]=0~,\quad [H^i(x),H^j(x)]=0,
\label{com}
\end{equation}
for all $x, i$ and $j$. Also, if $n$ is the number of nontrivial independent commuting matrices in a given family, the Hamiltonian $H$ can be written as 
\begin{equation}
H(x)=\sum\limits_{i=1}^{n} d_iH^i(x)~.
\label{lc}
\end{equation}
If $H(x)$ is written in the eigenbasis of $V$, $d_i$s are simply the eigenvalues of $V$.
The maximum value of $n$ can be $n_{max}=N-1$ and the corresponding family is called Type-1 with $H$ being completely integrable \cite{owusu2013, shastry2013, shastry2016, shastry2016a, shastry2005}.
 The number $n$ of conserved charges $H^i$ is, therefore, a measure of the strength of integrability in a system with Hamiltonian $H$.

In standard integrable systems, such as 1D Hubbard model and XXZ model, the conserved charges are polynomial in a tunable parameter \cite{shastry1986, grosse1989, shastry1986a, grabowski1995}.
An arbitrary linear superposition of all these charges will be another conserved charge and this charge will be an infinite power series in that parameter.

However, if the parameter, on which the conserved charges are dependent, is absorbed within the definition of the $T$ matrix (see Eq.~\ref{H}), the conserved charges are required to be constructed using the $T$ itself.
Analogous to the Type-1 Hamiltonian, the conserved charges $H^i$s are, therefore, formulated as infinite series:
\begin{equation}
H^i=P^i_0+P^i_1+P^i_2+\dots~,
\label{cc}
\end{equation}
with the coefﬁcients $P^i_m, (m\in [0,\infty))$ calculated using conditions of IMT (Eq.~\ref{com}).
We consider, $P^i_m$ is quadratic in the creation and annihilation operators , \textit{i.e.},
\begin{equation}
P^i_m=\sum_{jk}\eta_{jk}^{(m)}(i)c_k^\dagger c_j~,
\label{pim}
\end{equation}
where $\eta_{jk}^{(m)}(i)$ are determined in the form of a recursion relation (see Appendix~\ref{a1}), as follows:
\begin{equation}
\eta_{jk}^{(m+1)}(i)=\frac{1}{\epsilon_j-\epsilon_k}\sum_l[t_{jl}\eta_{lk}^{(m)}(i)-\eta_{jl}^{(m)}(i)t_{lk}]
\label{eta}
\end{equation}
for $j\neq k$. The $\epsilon_i$s are the eigenvalues of $V$ and $t_{ij}$s are the matrix elements of $T$ in the eigenbasis of $V$. We consider the diagonal term, \textit{i.e.}, $\eta_{jk}$ for $j= k$, to be zero for $m\geq 1$, because any other choices will not satisfy the condition Eq.~\ref{lc} \cite{modak2016}. This algorithm to determine the conserved charges for
the 1D Anderson-type ensemble is discussed in Ref.~\cite{modak2016}. The convergence of $\eta(i)$ with increasing $m$ implies the convergence of the series in the expression of $H^i$ in Eq.~\ref{cc}. This convergence confirms the existence of the corresponding conserved charge $H^i$. However, exact formulation of the $H^i$ is non-trivial. 
In this work, we will look for the number of the existing conserved charges, while tuning the inhomogeneity in the hopping of the Hamiltonian. Let us first define our system.

\section{The formulation of the model}
\label{s3}

A single particle system on a 1D lattice with random nearest-neighbor hopping is governed by a Hamiltonian whose matrix representation can be tridiagonal. A tridiagonal Hamiltonian matrix with random subdiagonals and superdiagonals reminds us of the matrix form of the well-known $\beta$-ensemble. The matrix representation of the $\beta$-ensemble is, in general, tridiagonal real symmetric where the primary diagonal elements are Gaussian distributed random numbers with mean zero and variance one, and the next diagonal elements are considered from a $\chi$ distribution with degrees of freedom $(N-i)\beta$. Here $i$ is the index of the diagonals and $\beta(>0)$ is generalized Dyson index. For $\beta\geq 1$, the subdiagonal and superdiagonal elements contribute significantly leading to a chaotic eigenspectrum.

In this work, we consider $V$ in Eq.~\ref{H} to be a $N\times N$ matrix representing the $\beta$-ensemble with $\beta = 1$ and the eigenvalues of $V$ thus follow the Wigner-Dyson statistics \cite{adway2022}. In its eigenbasis, $V$ becomes a diagonal matrix whose diagonal elements are its eigenvalues. This formulation is motivated by the requirements of IMT, as discussed in Sec.~\ref{s2}. In the same eigenbasis, we propose $T$ to be a $N\times N$ tridiagonal real symmetric matrix with zero diagonal elements. This form of $T$ will contribute only to the nearest neighbor hopping. Inhomogeneity in the hopping is introduced by considering the subdiagonals and superdiagonals of $T$ to be $\chi$-distributed random elements.
Therefore, the $N\times N$ Hamiltonian $H_{\textit{IH}}$ of our model consists of the following non-zero elements:
\begin{align}
&[H_{IH}]_{i,i}=\epsilon_i \quad \text{and}
\label{diag}\\
&[H_{IH}]_{i,i+1}=[H_{IH}]_{i+1,i}= t_{i,i+1} \nonumber\\
&~~~~~~~~~~~~~~~~~~~~~\text{with}\quad t_{i,i+1}\sim \frac{1}{\sqrt{2}}\chi_{(N-i)\beta}~,
\label{hop}
\end{align}
representing the inhomogeneous nearest neighbor hopping parameters and $\epsilon_i$s being the eigenvalues of $V$ matrix. 
This model from now on is referred to as \textit{inhomogeneous hopping (IH) random matrix ensemble}.

The parameter $\beta$ in the degrees of freedom of the $\chi$ distribution in the hopping terms (in Eq.~\ref{hop}) controls the deviation of $H_{IH}$ from its chaotic limit. We can, therefore, drop the parameter $x$ in Eq.~\ref{H} and tune $\beta$ to study the chaotic--integrable transition. In this model, $\beta$ is the parameter on which the conserved charges depend and it is thus convenient to construct conserved charges based on the hopping itself (Eq.~\ref{hop}).

The Hamiltonian of this matrix model in second quantization formulation
becomes quadratic:
\begin{equation}
H_{IH}=\sum_i \epsilon_i n_i + \sum_i t_{i,i+1}(c_i^\dagger c_{i+1} +h.c) 
\label{hbe}
\end{equation}
with $c_i^{\dagger}$ and $c_i$ being fermionic creation and annihilation operators, and $n_i=c_i^\dagger c_i$ is the number operator with $\epsilon_i$ being the onsite potential. Comparing with Eq.\ref{H}, we can write: $xT=\sum_i t_{i,i+1} (c_i^\dagger c_{i+1} +h.c)$ and $V=\sum_i \epsilon_i n_i$ with $i \in [0, (N-1))$.
Quadratic Hamiltonian $H_{IH}$ leads the conserved charges to be quadratic (see text before Eq.~\ref{com} in Sec.~\ref{s2}) and therefore validates the structure of the coefficients $P^i_m$ in Eq.~\ref{pim} required in the construction of conserved charges as discussed in Sec.~\ref{s2}.

Due to the distinct construction of this matrix model, its statistical properties are not expected to coincide with those of the $\beta$ ensemble. A detailed investigation of the statistical properties of this \textit{IH} random matrix ensemble is thus necessary.

\section{Statistical properties}
\label{s4}
In case of the $\beta$ ensemble, the Dyson index $\beta$ in the degrees of freedom of the $\chi$-distribution determines the significance of the subdiagonal and superdiagonal elements of the Hamiltonian matrix on its eigenspectrum. For $\beta \leq 1/N$, the contribution of such elements is negligible leading the eigenspectrum to show Poissonian statistics. On the other hand, $\beta \geq 1$ makes the spectrum chaotic. It is therefore inevitable to have an integrable to chaotic crossover for $1/N<\beta<1$ \cite{adway2022}.
Here we use similar logic to study such crossover for our \textit{IH} ensemble and hence tune $\beta$ in Eq.~\ref{hop} from $1/N \to 1$ using the following equation:
\begin{equation}
\beta=\frac{1}{N^{\gamma}}~,
\label{beta}
\end{equation}
where the parameter $\gamma \in \mathbb{R}$ is introduced for convenience.

\subsection{Eigenvalue Statistics}
\label{s41}
\begin{figure}[h!]
\centering
    \includegraphics[width=\linewidth]{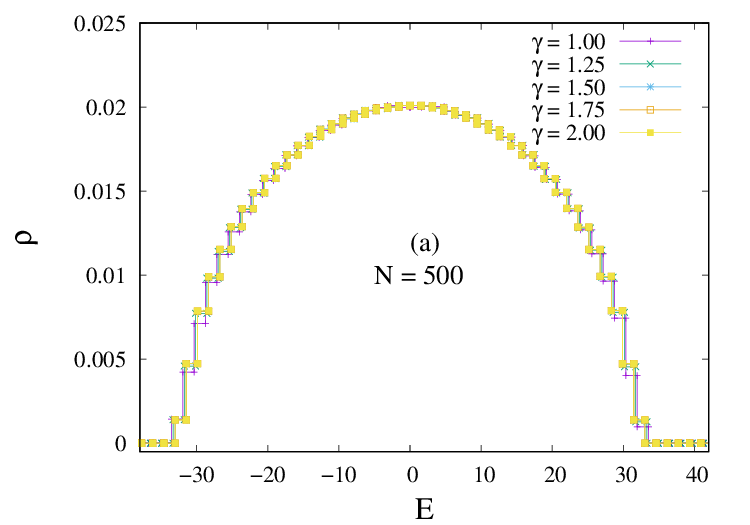}
	\includegraphics[width=\linewidth]{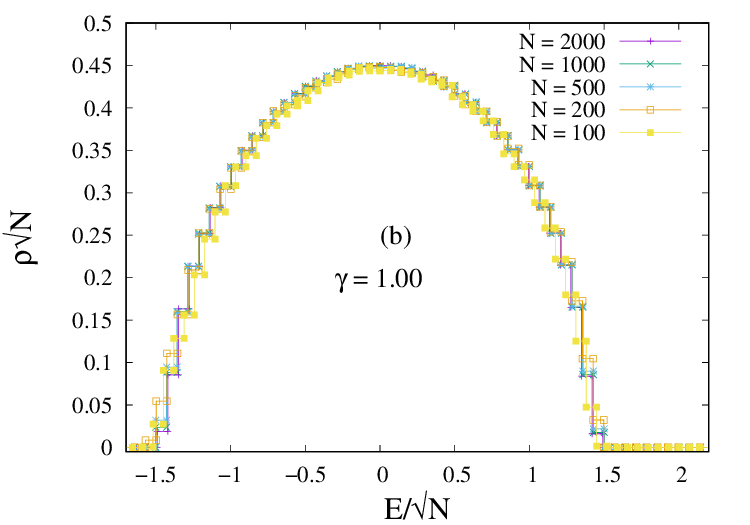}
	\caption{Density of states (a) for different values of $\gamma\geq 1$ for $5000$ samples of system size $N=500$. All the plots follow Wigner semicircle law. (b) shows the size dependence of density of states  at $\gamma=1$ for $5000$ samples of system sizes $N=100,200,500,1000$ and $2000$. If x-axis scales as $E\to \frac{E}{\sqrt{N}}$ and y-axis as $\rho\to\rho\sqrt{N}$, all the curves for different $N$ collapse onto a single curve.}
	\label{fig-lev}
\end{figure}

We study various spectral properties of our model by tuning the parameter $\gamma$ in Eq.~\ref{beta}. The Hamiltonian matrix $H_{IH}$ gradually becomes diagonal with increasing value of $\gamma$ from zero, as the off-diagonal elements become insignificant. This leads the density of states (DOS) of the \textit{IH} ensemble to Wigner semicircle at $\gamma\sim 1$ (Fig.~\ref{fig-lev}), unlike $\beta$-ensemble.
For $\gamma<1$, the DOS does not obey the semicircle law as it loses the symmetry of the eigenvalues around zero with decreasing $\gamma$ from one \footnote{Diagrams are not included to maintain clarity}.  However, the effect of higher values of $\gamma$ is insignificant on the density of states, as Fig.~\ref{fig-lev}(a) depicts the DOS superposed with one another for $\gamma>1$. 
To rule out the finite size effect on such behavior, we study the DOS at different values of $\gamma$ for different system sizes $N$. We plot Fig.~\ref{fig-lev}(b) for several $N$ for $\gamma=1$ only to maintain clarity. It reveals that the DOS for different sizes of Hamiltonian matrices perfectly overlap with each other at any value of $\gamma$, if the axes are scaled accordingly.

\begin{figure}[h!]
\centering
	\includegraphics[width=\linewidth]{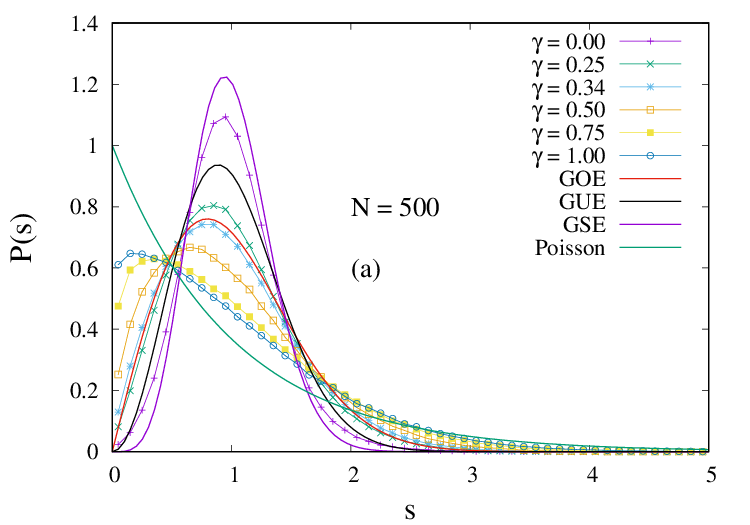}
	\includegraphics[width=\linewidth]{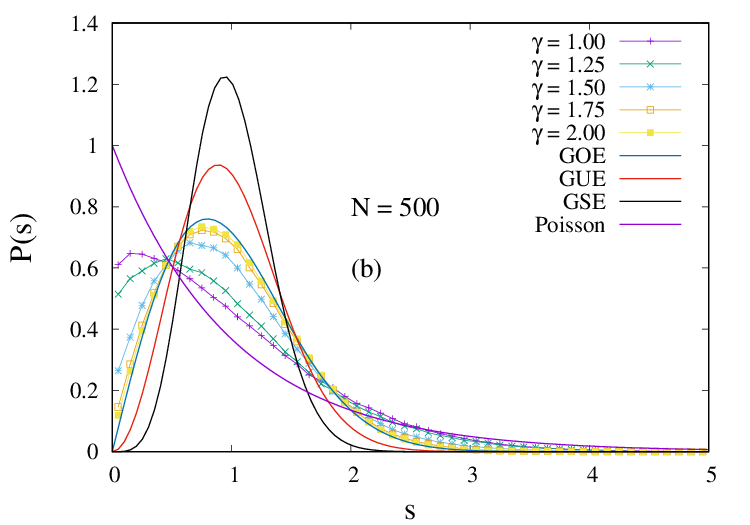}
	\caption{Nearest neighbor spacing distribution $P(s)$ for $5000$ samples of system size $N=500$. (a) depicts $P(s)$ for $0<\gamma<1$. It transits from GOE (at $\gamma\sim 0.34$) to near Poisson limit as $\gamma$ increases. (b) shows $P(s)$ for $\gamma>1$. It comes back towards GOE limit with a further increase of $\gamma$ and overlaps with Wigner distribution for $\gamma \sim  2$. $P(s)$ does not change any more if $\gamma$ increases further (not shown for clarity).}
	\label{fig-sp}
\end{figure}

Now, the question is how the spectral statistics look like for different values of $\gamma$?
The nearest neighbor spacing distribution $P(s)$ is a measure of short range correlations among the eigenvalues. It is essentially the probability distribution of finding two nearest neighbor levels at a distance $s$. The nature of $P(s)$, however, does not align with that of the DOS. As $\gamma$ increases from $0\to 1$, $P(s)$ goes from Wigner-Dyson to the Poisson statistics, though never superposes with the Poisson limit for finite size of $H_{IH}$ (Fig.~\ref{fig-sp} (a)). Both the plots in Fig.~\ref{fig-sp} exhibit that $P(s)$ behaves nearly integrable around $\sim \gamma=1$. Further increment of $\gamma$ (\textit{i.e.}, $\gamma>1$), unlike $\beta$-ensemble, moves the spacing distribution again towards Wigner-Dyson statistics and they almost overlap for $\gamma\gtrsim 2$ (Fig.~\ref{fig-sp}(b)). $P(s)$ remains there if $\gamma$ increases again. It is also notable in Fig.~\ref{fig-sp}(a) that $P(s)$ coincides with the GOE limit \footnote{GOE, GUE and GSE are the abbreviations of Gaussian Orthogonal Ensemble, Gaussian Unitary Ensemble and Gaussian Symplectic Ensemble. In random matrix theory, such ensembles show chaotic statistical behavior, often termed as Wigner-Dyson statistics, for ensembles of matrices with real (for GOE), complex (for GUE)  or quaternion (for GSE) elements. Their distribution is invariant under orthogonal, unitary or symplectic transformations respectively.} at $\gamma\sim 0.34$ for $H_{IH}$ of size $N=500$ for an ensemble of $5000$ matrices. If $\gamma$ decreases from this point, the spacing distribution approaches GUE statistics and a further reduction of $\gamma$ causes $P(s)$ to move towards GSE statistics for finite-sized system \cite{Note3}. We study other properties to investigate whether this behavior of $P(s)$ is due to the finite size effects.

\begin{figure}[h!]
\centering
	\includegraphics[width=\linewidth]{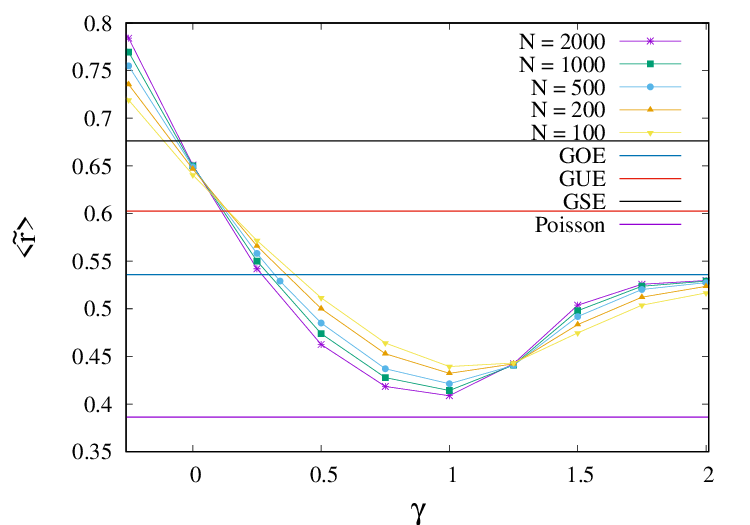}
	\caption{Size dependence of ensemble averaged ratio $\langle\tilde{r}\rangle$ of nearest neighbor spacing for $5000$ samples of system size $N=100, 200, 500, 1000, 2000$. As $\gamma$ increases from zero, it goes towards the Poisson limit touching GOE limit at $\gamma \simeq 0.29$ for $N=2000$ and $ \gamma \simeq 0.42$ for $N = 100$. $\gamma=0$ is a critical point with $\langle\tilde{r}\rangle = 0.65$ lying the statistics between GUE and GSE limits. $\langle\tilde{r}\rangle$ attains maximum integrability at $\gamma\simeq 1$ for systems of all sizes $N$. But larger ones become more integrable with $\langle\tilde{r}\rangle\to$ Poisson limit. It again moves toward GOE statistics and superposes with it around $\gamma\gtrsim 2$. Here also larger systems attain at smaller values of $\gamma$. All the curves again cross with each other at $\gamma\simeq 1.25$ with $\langle\tilde{r}\rangle=0.44$.}
	\label{fig-rav}
\end{figure}

Nearest neighbor spacing distribution $P(s)$ requires \textit{unfolding} of the eigenspectrum. The ratio of the nearest neighbor spacing is usually  studied to minimize any errors that may arise due to unfolding. This is possible because spacing ratio $r$ does not require the spectrum to be unfolded. The ratio of spacing between two consecutive nearest neighbor levels $e_i$ and $e_{i+1}$ is defined as $r_i=s_{i+1}/s_i$ with $s_i=e_{i+1}-e_i$  \cite{atas2013}.
The spacing ratio is distributed as $P(\tilde{r})=\sum_{i=1}^N\langle\delta (\tilde{r}-\tilde{r}_i)\rangle$ with $\tilde{r}_i={\rm min}\Big(r_i, \frac{1}{r_i}\Big)$.
The distribution $P(\tilde{r})$ exactly verifies the result we found in Fig.~\ref{fig-sp} \cite{Note2}.
The ensemble average of the spacing ratio $\langle\tilde{r}\rangle$ can measure more accurately how close the spectrum is to the Poisson or Wigner limit at different values of $\gamma$, as $\langle\tilde{r}\rangle$ narrows the result down to a single point. Fig.~\ref{fig-rav} perfectly verifies the outcomes we obtained by studying $P(s)$ (Fig.~\ref{fig-sp}) and $P(\tilde{r})$, \textit{i.e.}, with increasing $\gamma$ from its zero value, $\langle\tilde{r}\rangle$ approaches to the Poissonian limit, although it does not completely attain the limit for a finite-sized system. The size dependence in Fig.~\ref{fig-rav} also shows that increasing system size $N$ leads the system towards more integrability at $\gamma=1$, suggesting complete integrability in the limit $N\to \infty$. Again, $\langle\tilde{r}\rangle$ behaves exactly GOE-like at $\gamma\sim 0.34$ for $N = 500$ (with ensemble of $5000$ matrices), analogous to Fig~\ref{fig-sp}(a). For larger system, the value of $\gamma$, at which $\langle\tilde{r}\rangle$ overlaps with its GOE-value, is closer to zero. Below that point, $\langle\tilde{r}\rangle$ goes beyond GOE statistics. 
At $\gamma\simeq 0, \langle\tilde{r}\rangle$ for different system sizes intersect at a critical point lying between GUE and GSE statistics with $\langle\tilde{r}\rangle = 0.65$.
It is again shown in Fig. \ref{fig-rav} that $\langle\tilde{r}\rangle$ tends to return from near-Poisson to the Wigner distribution as $\gamma>1$ and coincides with the GOE statistics at $\gamma\gtrsim 2$ for any system size. Here also, the larger system approaches to the GOE-limit faster. 
After crossing the integrable limit at $\gamma\simeq 1$, with the increasing value of $\gamma$, all the curves for different system sizes again superpose with each other around the critical point $\gamma\simeq 1.25$ with $\langle\tilde{r}\rangle=0.44$.

\subsection{Eigenfunction Statistics}
\label{s42}

The inverse participation ratio (IPR) denotes the inverse of the effective number of basis states occupied by an eigenstate.
When an eigenstate is uniformly spread over all the basis states, corresponding IPR is $1/N$, implying complete delocalization in Wigner-Dyson limit. In contrast, if the number of basis states contributing to an eigenstate decreases, its IPR increases and reaches unity when only a single basis state remains. This corresponds to complete localization in Poissonian limit.
In case of our system, we find localization in the eigenstates increases with $\gamma$. For $\gamma<1$, both the eigenvalues and their IPRs are not symmetric around zero. Although the number of negative eigenvalues becomes same as that of positive eigenvalues for $\gamma>1$, their corresponding IPRs never achieve such symmetry. We also find that, for any value of $\gamma$, as the positive eigenvalues move away from their bulk, localization in their eigenstates increases. The IPR corresponding to the largest eigenvalue reaches its maximum value (\textit{i.e.}, IPR $=1$) for any $\gamma\geq 0.34$ for $N=500$ (studied for $5000$ samples) \cite{Note2}. Difficulty to analyse such IPR curves directs us to explore other eigenvector properties.

The fractal dimension $D_q$ (for $q\in[2,\infty)$) is another measure of eigenstate localization. For a finite-sized system, it is defined as $D_q^{(N)}=-(q-1)^{-1}\langle {\rm ln} I_q\rangle/{\rm ln} N$, with $I_q$ being the $q^{\rm th}$ moment of local intensity. If $D_q^{(N)}=1$, the eigenstates are completely extended and $D_q^{(N)}=0$ confirms exact localization of them. For $0<D_q^{(N)}<1$, eigenstates are fractal or multifractal. For system size $N\to \infty$, $D_q^{(N)}$ becomes $D_q$. The IPR is essentially is the $2^{\rm nd}$ moment ($q=2$), denoted as $I_2$. Corresponding fractal dimension $D_2^{(N)}$ is considered to be a better measure of eigenstate localization compared to the $I_2$, as the information here is confined to a single point.
\begin{figure}[h!]
\centering
	\includegraphics[width=\linewidth]{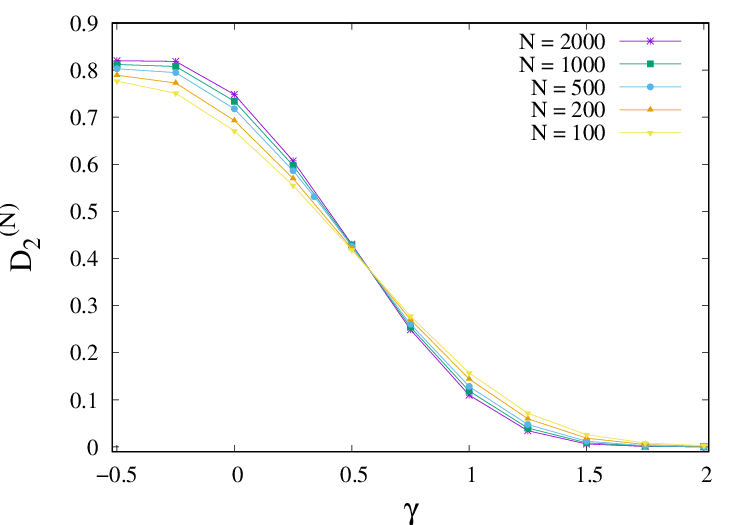}
	\caption{Size dependence of the fractal dimension $D_2^{(N)}$ for $5000$ samples of system size $N=100,200,500,1000,2000$. At $\gamma=-0.25$, the system is maximally delocalized with $D_2^{(N)}\to 1$. At this point, delocalization increases, as $N$ increases. The eigenstates become multifractal with increasing $\gamma$ and they reach a complete localization $D_2^{(N)}=0$ around $\gamma\simeq 2$. As the system size $N$ increases, localisation is attained at a smaller value of $\gamma$. With a further increment of $\gamma$, the eigenstates remain localized. All the curves for any system size intersect at $\gamma\simeq 0.57$ with $D_2^{(N)}\simeq 0.38$.}
	\label{fig-d2}
\end{figure}
We plot Fig.~\ref{fig-d2} for $q=2$ only. It shows that with increasing $\gamma$, the localization in eigenstates increases and they become completely localized around $\gamma\simeq 2$ for any system size. The size dependence in Fig.~\ref{fig-d2} depicts that eigenstates of larger systems localize at smaller $\gamma$, suggesting that for a system with $N\to \infty$, localization ($D_2=0$) is attained at $\gamma\to 1$. The eigenstates remain localized for further increment of $\gamma$, \textit{i.e.}, $D_2^{(N)}$ never goes again towards the Wigner-Dyson limit, unlike the spectral properties of our \textit{IH} ensemble. The size dependence in Fig.~\ref{fig-d2} also indicates that increasing system size $N$ leads $D_2^{(N)}$ towards GOE-limit at $\gamma=0$. The value of $D_2^{(N)}$ maximizes ($\sim 1$) at $\gamma = - 0.25$ for any size of the system. For a fixed system size, no further decrease in $\gamma$ makes the system more chaotic. It is also notable that $D_2^{(N)}\simeq 0.38$ at the critical point $\gamma\simeq 0.57$, where all the curves for any system size coincide. This critical point is not consistent with that of the average spacing ratio $\langle\tilde{r}\rangle$ at $\gamma\simeq 1.25$. 
We therefore explore the criticality by studying the fractal dimensions for higher orders ($q>2$) and observe that the value of $\gamma$ at the critical point is highly dependent on the order $q$ \cite{Note2}. This implies that eigenstates with $0<D_q^{(N)}<1$ are multifractal in nature. We also notice that $\gamma$ at the critical point increases with $q$ and predict that it reaches $\gamma\simeq 1.25$ at $q\to \infty$.

As we know, the integrability in a system is associated with Poissonian spectrum of eigenvalues and localization in eigenstates leads to the Poissonian eigenspectrum; we therefore study both the eigenvalue and eigenvector statistics here to explore the integrability in our system. We attempt to verify the accuracy of all these statistical measures using IMT in the following sections. However, before that, we compare the statistical properties of our system with those of the $\beta$ ensemble.

\subsection{Comparison with $\beta$-ensemble}
\label{s53}

The statistical properties of the well-known $\beta$-ensemble show a clear crossover from Wigner-Dyson to Poissonian statistics as the tuning parameter $\gamma$ in $\beta$ (Eq.~\ref{beta}) increases from $0\to 1$ \cite{adway2022}. The diagonal elements in the Hamiltonian matrix of $\beta$-ensemble are normal distributed random numbers ($\sim \mathcal{N}(0,1)$). At higher value of $\gamma~(\sim 1)$, the subdiagonals and superdiagonals become insignificant, driving the system to an integrable one by effectively making the Hamiltonian a diagonal matrix. This phenomenon is not valid for our system as the diagonals in the Hamiltonian matrix $H_{IH}$ in Eq.~\ref{hbe} are the eigenvalues following GOE statistics
and therefore, at $\gamma=1$, the density of states of our system becomes semicircular.
However, the spectral properties like spacing distribution $P(s)$ and the average ratio $\langle\tilde{r}\rangle$ of spacing follow the trend similar to those of the $\beta$-ensemble for $\gamma \leq 1$ (Fig.~\ref{fig-sp}(a)
and \ref{fig-rav}). The spectrum in case of the $\beta$-ensemble though remains Poissonian for $\gamma>1$, Fig.~\ref{fig-sp}(b) and \ref{fig-rav} exhibit again a smooth cross-over towards the Wigner-Dyson statistics in case of our \textit{IH} ensemble. 
One of the important measures of the eigenstate dynamics, the fractal dimension $D_2$, shows a crossover from higher value ($\sim 1$) to zero with increasing $\gamma$. Similar crossover was found in case of the $\beta$-ensemble which is analogous to its spectral transition \cite{adway2022}. However, in this model, $D_2^{(N)}$ does not reach to the value one at $\gamma = 0$ for finite-sized systems (Fig.~\ref{fig-d2}), unlike $\beta$-ensemble. Also, $D_2^{(N)}$ here goes to zero at $\gamma > 1$, not at $\gamma = 1$ like $\beta$-ensemble. This behavior of eigenstates is also not consistent with the eigenvalue statistics of our \textit{IH} ensemble. We therefore take the route of IMT to seek such consistency. In the next section, we try to estimate the conserved charges and consequently determine the integrability in our system with $\gamma$ as a tuning parameter.

\section{Estimation of conserved charges}
\label{s5}
The conserved charges $H^i$ described in Sec.~\ref{s2} is dependent on the hopping elements $t_{ij}$ of $H_{IH}$ (see Eq.s~\ref{cc}, \ref{pim} and \ref{eta}). The tunable parameter $\gamma$ (Eq.~\ref{beta}) controls those hopping elements microscopically (Eq.~\ref{hop}). In this section, we predict the behavior of $H^i$ by varying $\gamma$. How the convergence of $\eta$ in Eq.~\ref{eta} depends on $\gamma$ is the primary focus of this section, as its convergence confirms the existence of corresponding conserved charges. In addition, the number of conserved charges as a function of $\gamma$ is another important quantity that we investigate, as this number is expected to vary with the strength of integrability. For a finite-sized system, we know that the number of conserved charges attains the value $N-1$ in the completely integrable limit.

First, we numerically calculate  $\eta (i=0)$ in Eq.~\ref{eta} for a certain conserved charge $H^0$ at several values of $m$ by tuning $\gamma$ in hopping parameters.
\begin{figure}[h!]
\centering
	\includegraphics[width=\linewidth]{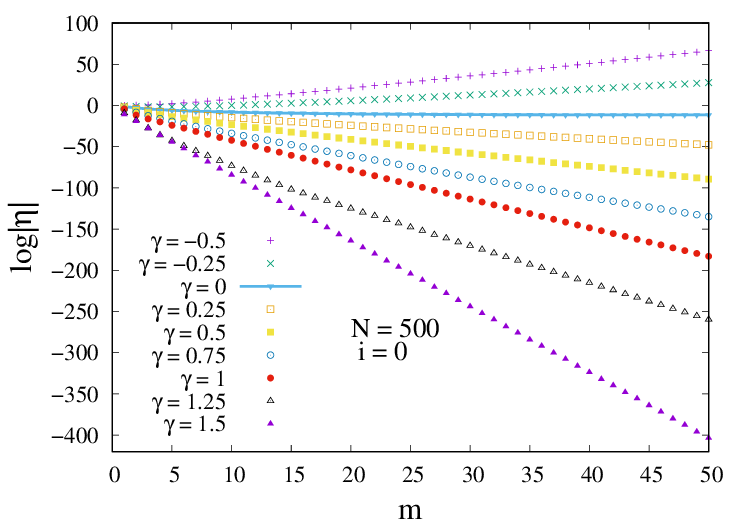}
	\caption{$\text{log}|\eta|$ with increasing $m$ for conserved charge $H^0$ for system size $N=500$. $\eta$ starts to converge at $\gamma=0$ and with increasing $\gamma$, the convergence becomes faster.}
	\label{fig-eta}
\end{figure}
Figure~\ref{fig-eta} shows that when $\gamma = 0$, $\eta$ starts to converge, ensuring that $H^0$ exists. Existence of the conserved charges implies breaking of the ergodicity of the system; \textit{i.e.}, the eigenstates of the system no longer remain completely delocalized.
As $\gamma$ increases, $\eta$ converges faster and thus, $\text{log}|\eta|$ falls in sharper slopes with power $m$. Quicker convergence of $\eta$ with increasing $\gamma$ resembles with stronger localization if compared with the eigenstate properties of the system discussed in Sec.~\ref{s42}. In general, it is found that localized eigenfunctions correspond to the eigenspectrum exhibiting Poissonian statistics. Our system shows that the short range correlation measures of eigenvalues leave their chaotic limit for $\gamma>0$ for finite-sized system (Sec.~\ref{s41}). The eigenvalue statistics then move towards the integrable limit with increasing $\gamma$ which is again comparable with faster convergence of $\eta$.

If $\gamma$ increases further ($\gamma>1$), slopes of $\text{log}|\eta|$ vs $m$ become sharper  only to confirm near-to-complete localization of the states. This behavior is, though, compatible with the eigenfunction statistics (see the fractal dimension $D_2^{(N)}$ shown in Fig.~\ref{fig-d2}), not consistent with the eigenvalue statistics. Fig.~\ref{fig-sp}(b) and \ref{fig-rav} show that the spectrum of our \textit{IH} model starts going back towards Wigner-Dyson statistics as $\gamma>1$. However, for such higher $\gamma$'s, the calculation of $\eta$ becomes quite difficult even for small values of $m$, as the hopping terms $t_{ij}$ (Eq.~\ref{hop}) in the expression of $\eta$ (Eq.~\ref{eta}) tend to zero and consequently, $\text{log}|\eta|$ becomes undefinable. Here, we investigate more characteristics of $\eta$ to validate all of these findings.

\begin{figure}[h!]
\centering
	\includegraphics[width=\linewidth]{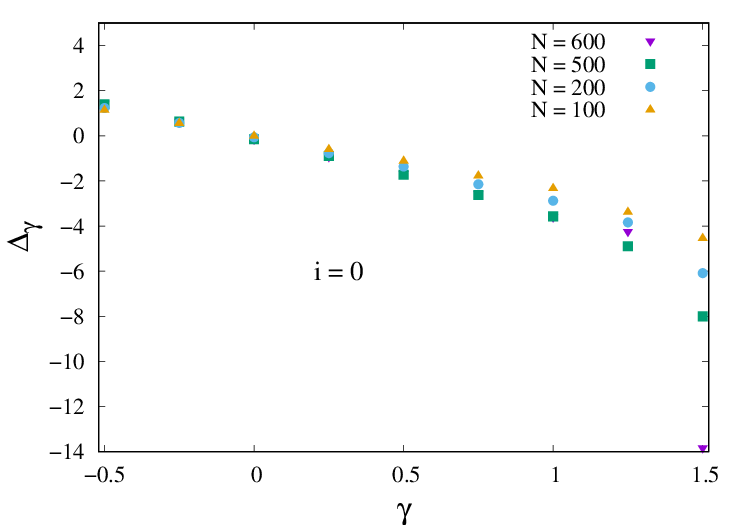}
	\caption{Size dependence of slope $\Delta_{\gamma}$ with respect to $\gamma$ for system-size $N=100, 200, 500, 600$. The $|\Delta_{\gamma}|$ is higher for the larger system for the same $\gamma$. This difference is more prominent at higher $\gamma$. All the curves break ergodicity $\gamma\sim 0$.}
	\label{fig-slope-size}
\end{figure}

The $\text{log}|\eta|$ curves being straight line with respect to $m$ immediately prompt to calculate their slopes $\Delta_{\gamma}$. It is obvious to get negative $\Delta_{\gamma}$s when $\gamma\geq 0$ and $|\Delta_{\gamma}|$ increases as $\gamma$ increases (Fig.~\ref{fig-slope-size}). The finite size effect in IMT is also depicted in Fig.~\ref{fig-slope-size}. The larger system has higher $|\Delta_{\gamma}|$ for the same $\gamma$; and as $\gamma$ increases from zero, finite size effect becomes more prominent, leading the $|\Delta_{\gamma}|$ to a significantly higher value for larger system. That is how the system size $N$ affects localization and thus integrability in the system. 

The conserved charge has been discussed so far for $i=0$. It is necessary to know if there exist other conserved charges. We understand from Fig.~\ref{fig-eta} that convergence of $\eta$ leads to negative slope $\Delta_{\gamma}$ of $\text{log}|\eta|$ vs $m$ curve. The negative value of $\Delta_{\gamma}$ implies the existence of corresponding conserved charge. A completely integrable system of size $N$ is expected to have $N-1$ conserved charges. This number reduces as the system goes towards chaotic limit and it reaches zero when the system becomes completely chaotic. We calculate the slopes of $\text{log}|\eta(i)|$ vs $m$ curves for $0\leq i<(N-1)$ and count the number $n$ of negative slopes as a function of $\gamma$. Therefore, $n$ becomes the number of existing conserved charges.
\begin{figure}[h!]
\centering
	\includegraphics[width=\linewidth]{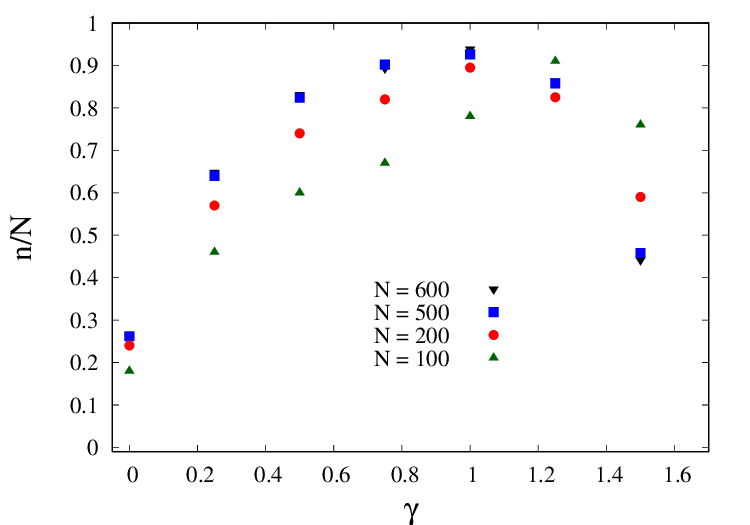}
	\caption{Scaled number of conserved charges calculated from $\Delta_{\gamma}$s across $\gamma$ for system sizes $N=100,200,500,600$. $n$ increases with $\gamma$ for $0<\gamma<1$ and at $\gamma\sim 1$, it becomes maximum $n\simeq N$. Again, for $\gamma>1$, it decreases rapidly though actual number of conserved charges remains equal to $N$ (not shown here). Also, for the same $\gamma$, larger system (high $N)$ have higher value of $n$ compared to the smaller ones.}
	\label{fig-number}
\end{figure}
In Fig.~\ref{fig-number}, the number $n$, scaled by the system size $N$, is plotted for different $\gamma$ values for several system sizes. It is clear that the number $n$ of conserved charges increases as $\gamma$ goes from $0\to 1$, implying a crossover from near-chaotic to near-integrability of our finite-sized \textit{IH} model with increasing $\gamma$. This result is consistent with the statistical properties of our system described in Sec.~\ref{s4}.
However, for $\gamma>1$, Fig.~\ref{fig-number} shows that the number of conserved charges reduces! This behavior initially seems to align with the eigenvalue statistics of the \textit{IH} ensemble (Sec.~\ref{s41}). But with increasing $\gamma$,
the hopping terms $t_{ij}$ in $H_{IH}$ (Eq.~\ref{hbe}) become insignificant leading the Hamiltonian to be a diagonal matrix. The calculation of conserved charges described in Sec.~\ref{s2} is valid for a Hamiltonian of the form Eq.~\ref{H}. For diagonal $H_{IH}$, $\eta$ is just equal to $\eta^{(0)} = 1$ (see Eq.~\ref{eta00} in Appendix~\ref{a1}). This prompts that the number of conserved charges to be equal to the system size. However,  Fig.~\ref{fig-number} only indicates those conserved charges which are calculated from the slopes $\Delta_{\gamma}$.
 Fig.~\ref{fig-number} also depicts that there is a significant effect of finite system size $N$ on the number $n$ of conserved charges even after scaling. A system with larger size $N$ has larger number $n$ of conserved charges for the same value of $\gamma$ compared to the smaller sized system. The scaled number $n/N$ of conserved charges never reaches the value one if they are calculated using slopes $\Delta_{\gamma}$ for a finite system but becomes $\simeq 1$ around $\gamma\sim 1$. 
We discuss all the above results we found in the following section.

\section{Conclusions and discussions}
\label{s6} 

We study a particular type of 1D random matrix ensemble with inhomogeneous hopping and chaotic primary diagonal, called \textit{IH} ensemble, to explore a crossover from chaotic to integrable limit by tuning the hopping elements. This crossover is studied here in terms of conserved charges. 
The estimation of the conserved charges was introduced to explain the localization in a 1D Anderson-type ensemble in Ref.~\cite{modak2016}. 
This work motivated us to study the localization in a system which is different from the 1D Anderson ensemble with hopping being inhomogeneous. 
The localization described here goes hand-in-hand with the integrability in the system, as conserved charges are estimated using IMT. In this work, we are interested to explain a $\beta$-dependent chaotic--integrable crossover of the \textit{IH} model, introduced in Sec.~\ref{s3}, in terms of the conserved charges. This kind of crossover was previously found in case of the $\beta$ ensemble, termed as non-ergodic-extended phase \cite{adway2022}.

The tuning parameter $\gamma$ (Eq.~\ref{beta}) in our model controls the convergence of $\eta$ (Eq.~\ref{eta}) in the expression of conserved charges (discussed in detail in Sec.~\ref{s2} and \ref{s3}). Increasing $\gamma$ from its zero value increases the rate of convergence (Fig.~\ref{fig-eta}). In the range $0<\gamma\leq 1$, eigenvalue statistics along with eigenstate dynamics show an intermediate statistics between the Wigner-Dyson and Poissonian limits. We explain this intermediate behaviour in terms of conserved charges based on the slope of $\text{log}|\eta|$ vs $m$ curve in Fig.~\ref{fig-eta}. We have also studied the dependence of the slopes (and consequently the conserved charges) on the system sizes, verifying the finite size effect on IMT (Fig.~\ref{fig-slope-size}). The number of conserved charges $n$ are calculated from the slopes and it increases as the system goes towards the Poisson limit (Fig.~\ref{fig-number}). A completely integrable system of size $N$ depicting Poisson statistics has number of conserved charges $n\equiv n_{max} = N-1$. Figure~\ref{fig-number} shows that $n/N$ does not attain the value one for a finite-sized system of our \textit{IH} ensemble, \textit{i.e.}, it never becomes completely integrable at $\gamma\leq 1$. 
 The eigenvalue statistics of the \textit{IH} ensemble in Sec.~\ref{s41} also show a finite size effect as it does not reach completely integrable limit at any value of $\gamma$.

For $\gamma>1$, the eigenspectrum starts moving towards the Wigner-Dyson statistics again but the dynamics of the eigenstates keep moving towards integrability, not following the eigenvalues (Sec.~\ref{s41} and \ref{s42}). The construction of the Hamiltonian matrix (Eq.~\ref{H} and \ref{hbe}) of the system in IMT is the reason for this behavior. For higher $\gamma$, the off-diagonal elements of the Hamiltonian matrix $H_{IH}$ become insignificant, which leads its statistical properties to solely depend on the diagonals. The diagonal elements are the eigenvalues obeying GOE statistics, satisfying the conditions of IMT (Sec.~\ref{s2}). This is why the spectrum of the Hamiltonian again goes towards the chaotic limit for higher values of $\gamma$. However, the eigenvectors of a diagonal matrix is always localized, independent of the values of the matrix elements. The eigenstate dynamics of $H_{IH}$, therefore, never come back again towards Wigner-Dyson limit; it becomes more and more localized with higher $\gamma$ and saturates when completely localized. The convergence of $\eta$ for $\gamma>1$ initially becomes faster (Fig.~\ref{fig-eta}) confirming more localization;  but the hopping terms $t_{ij}$ in the recursion relation (Eq.~\ref{eta}) tend to zero as $\gamma$ increases further. 
Thus, as the Hamiltonian $H_{IH}$ becomes diagonal, $\eta$ can no longer be calculated following the process described in Sec.~\ref{s2}. For diagonal $H_{IH}$, corresponding conserved charges solely depend on $\eta^{(0)}$. This leads number of conserved charges to be exactly equal to the system size, confirming complete integrability in our system at $\gamma>>1$.

However, our interest limits the $\gamma$ in the range $(0,1)$, where statistics are intermediate showing a crossover from chaotic to integrable spectrum with increasing $\gamma$ for \textit{IH} random matrix ensemble. We have seen that this kind of transition is so far explained in literature in terms of the random matrix properties of the ensemble. To the best of our knowledge, properties and number of conserved charges are used for the first time to explain such transition, though the exact formulation of the conserved charges is yet to be determined. The \textit{IH} ensemble we introduced by incorporating inhomogeneity in the hopping of a 1D Anderson ensemble has turned into an appropriate choice for studying delocalization to localization transition in terms of conserved charges.

\section*{Acknowledgement} 
The author gratefully acknowledges Dr. Anandamohan Ghosh for his valuable feedback and suggestions.

\appendix*
\section{}
\label{a1}
 
The commutation relations \eqref{com} provide the following expression if we consider the definition of $H$ from Eq.~\ref{H} and $H^i$ from Eq.~\ref{cc}:
\begin{align}
[P^i_0,V]&+\sum_{m=0}([P^i_m,T]+[P^i_{(m+1)},V])=0.
\label{pvpt}
\end{align}
Due to linear independence, 
\begin{align}
[P^i_0,V]&=0
\label{pvpt}\\
{\rm and}~~~~~
[P^i_m,T]+[P^i_{(m+1)},V]&=0~~.
\label{p-com}
\end{align}
Using $V=\sum_i \epsilon_i n_i$ and the definition of $P^i_m$ in Eq.~\ref{pim}  for $m=0$, one can find 
\begin{equation}
\eta_{ab}^{(0)}(i)=\delta_{ia}\delta_{ib}~~.
\label{eta00}
\end{equation}
This is the initial condition to solve $\eta$ in Eq.~\ref{eta}. 
$\eta_{ab}^{(0)}(i)$ directly leads to the conserved charges of diagonal $V$ which is nothing but the number operators $n_i$. 

From Eq.\ref{p-com} , we can write
\begin{align}
[P^i_{(m+1)},V]=-[P^i_m,T]~~.
\label{p-com2}
\end{align}
Using the anti-commutation relations of the creation and annihilation operators, LHS of Eq.~\ref{p-com2} becomes 
\begin{align}
 &\left[P^i_{m+1},V\right]\nonumber\\
  &=\sum\limits_{ab}\eta_{ab}^{(m+1)}(i)\left(\epsilon_b-\epsilon_a\right)c_b^{\dagger}c_a
 \label{lhs}
\end{align}
and RHS turns out to be 
\begin{align}
& -\left[P^i_m,T\right]\nonumber\\
&=\sum\limits_{ab}c_b^{\dagger}c_a\sum\limits_{j}\left[t_{aj}\eta_{j b}^{(m)}(i)-t_{jb}\eta_{a j}^{(m)}(i)\right]
\label{rhs}
\end{align}
after a detailed analysis.

Comparing the coefficients of $c_b^{\dagger}c_a$ in both the LHS~\eqref{lhs} and RHS~\eqref{rhs}, we have 
\begin{align}
 \eta_{ab}^{(m+1)}(i)&=\frac{1}{\epsilon_b-\epsilon_a}\sum\limits_{j}\left[t_{aj}\eta_{j b}^{(m)}(i)-t_{jb}\eta_{a j}^{(m)}(i)\right]~.
\end{align}
 for $a\ne b$. This is Eq.~\ref{eta} in the main text.

\bibliographystyle{unsrt}
\bibliography{newb}
\end{document}